\documentclass{article}
\usepackage[T1]{fontenc} 
\usepackage[utf8]{inputenc} 
\usepackage{ismir,amsmath,cite,url}
\usepackage{graphicx}
\usepackage{color}

\usepackage{lineno}
\usepackage{amsmath}
\usepackage{amsthm}

\title{Detecting music deepfakes is easy but actually hard}

\threeauthors
  {Darius Afchar} {Deezer Research}
  {Gabriel Meseguer-Brocal} {Deezer Research}
  {Romain Hennequin}{{\tt research@deezer.com}}

\def\authorname{D. Afchar, R. Hennequin, and G. Meseguer-Brocal}
\newcommand{\ie}{\textit{i.\,e.,} }
\newcommand{\eg}{\textit{e.\,g.,} }

\begin{document}

\maketitle

\begin{abstract}
In the face of a new era of generative models, the detection of artificially generated content has become a matter of utmost importance. The ability to create credible minute-long music deepfakes in a few seconds on user-friendly platforms poses a real threat of fraud on streaming services and unfair competition to human artists. This paper demonstrates the possibility (and surprising ease) of training classifiers on datasets comprising real audio and fake reconstructions, achieving a convincing accuracy of \textbf{99.8\%}. To our knowledge, this marks the first publication of a music deepfake detector, a tool that will help in the regulation of music forgery.
Nevertheless, informed by decades of literature on forgery detection in other fields, we stress that a good test score is not the end of the story. We step back from the straightforward ML framework and expose many facets that could be problematic with such a deployed detector: calibration, robustness to audio manipulation, generalisation to unseen models, interpretability and possibility for recourse. This second part acts as a position for future research steps in the field and a caveat to a flourishing market of fake content checkers.
\end{abstract}

\section{Introduction}

Generative models have gained tremendous popularity in the past couple of years. One concern surrounding their use is detecting generated content within a body of genuine human-made content. Within the framing of forgery, such artificial generations are commonly referred to as \textit{deepfakes}. The first evidence of deepfakes regarded facial replacement in videos in 2017, but since then, the term has acquired a broader meaning to designate fake images, illustrations, text, audio, and music. Compared to more traditional manual editions (\eg photoshopping an image), deep learning's generations usually imply that very realistic content can be generated in a matter of seconds and on a large scale. More broadly, discussion around deepfakes has regarded political risks (\eg Zelensky's deepfake \cite{url_zelensky}), scams (\eg voice spoofing for fake ransoms\cite{url_scam}), sexual harassment and violence (\eg pornography \cite{url_pornography}), unfair competition against artists (\eg recent Hollywood strike \cite{url_strike}). As for generative models, many questions have been raised about their reliability, monopoly by under-regulated big tech actors, ecological and social impact \cite{bender2021parrot, gebru2023artists, url_finetuning, gautam2024melting, crawford2021atlas}.

In 2023, a new sudden wave of generative models has rendered the risk of \textit{music} deepfakes more sensible than before \cite{schneider2023mousai, agostinelli2023musiclm, huang2023noise2music, yang2023diffsound, garcia2023vampnet, copet2024simple, chen2024musicldm}. Several user-friendly services have also recently emerged and democratised the creation and diffusion of artificial music (\eg \cite{riffusion, suno, stability_audio, udio}).
Music deepfakes now pose a growing problem for music artists and labels. If not properly detected, they could lead to significant copyrighting and fraud issues. This underscores the urgency of developing effective fake music detection methods.

While studies have been conducted on detecting fake audio and singing voices \cite{wu2017asvspoof, zang2024singfake}, we present a novel approach in this paper. We propose the first general-purpose music deepfake detector, a significant advancement that also includes generated instrumental parts. Our focus is on the trending waveform generators mentioned earlier. We leave symbolic or MIDI-based synthesis models for future exploration.
Using somewhat straightforward convolutional models, we show that almost perfect detection scores (>99\% accuracy) are easy to obtain.

Although music deepfake detection is novel, we do not conduct our research in a vacuum. The first models for video deepfake detection were published in 2018 \cite{afchar2018mesonet, rossler2019faceforensics}. While these detectors are not directly transferable to the specifics of music, we can at least anticipate having to deal with similar subtleties and research questions raised in this literature \cite{mirsky2021creation, lin2024detecting}. Therefore, in the second part of our paper, we take a step back on our seemingly impressive results and expose many caveats to deepfake detection: calibration issues, robustness to audio manipulation, interpretability for recourse, and generalisation to unseen generators. In a nutshell, \textit{we have to look beyond performance scores}, no matter how good they look.

This paper serves as a first research study on music deepfakes and a proof of concept that they can be detected, but also as a position and message for the research community on the many facets and challenges beyond the usual ML framework of training and testing a model that we urge to consider for the future research steps of this topic.

Our code is available at \url{github.com/deezer/deepfake-detector}.

\section{Detecting fake music seems easy}

There are many ways to tackle music deepfake detection. In this section, we first discuss our choice of framework and its advantage to solve the task, the employed data as well as some first surprisingly convincing detection scores.

\subsection{Proposed framework}

Motivated by the democratisation of online service that can generate minutes-long synthetic music, we restrict the scope of our paper to waveform based generators, common in these services: \eg WaveNet \cite{van2016wavenet}, MelGAN \cite{kumar2019melgan}, HiFiGAN \cite{kong2020hifi}, Jukebox \cite{dhariwal2020jukebox}, Musika! \cite{pasini2022musika}, Moûsai \cite{schneider2023mousai}, MusicLM \cite{agostinelli2023musiclm}, Noise2Music \cite{huang2023noise2music}, DiffSound \cite{yang2023diffsound}, VampNet \cite{garcia2023vampnet}, MusicGen \cite{copet2024simple},
MusicLDM \cite{chen2024musicldm}.
This list is not exhaustive of the swarm of published music generation methods, especially in 2023. We only provide a representative subset.

While it is impossible to account for all particularities, we can usually break down these models into two parts. First, an autoencoder (AE) is trained to compress bits of raw audio into an easier representation to process and to invert this representation into an audio signal (\ie vocoder).
For instance, mel-spectrogram representations were often used (\eg \cite{kumar2019melgan}) before being replaced by more recent so-called neural codecs -- as Soundstream \cite{zeghidour2021soundstream}, Encodec \cite{defossez2022high} or DAC \cite{kumar2024high} -- that demonstrated better reconstructions. For interested readers, the latter commonly employ discretised latent spaces as codebooks of tokens -- \eg \textit{Residual Vector Quantization} (RVQ) \cite{razavi2019generating}. Then, a second internal module is usually trained to learn to continue the compressed sequence temporally or generate it conditioned on text inputs, depending on the considered task. For instance, large-language-model (LLM) inspired architectures have been proposed for the role \cite{garcia2023vampnet}.
In layperson's terms, we can summarise that the AE does the waveform synthesis part while the LLM does the semantic work of generating a coherent musical sequence through time.

Detecting that a music sequence was artificially generated can be tricky. With the risk of falling into anthropomorphism, this equates to trying to learn a musician's style: \eg MusicGen might always interpret a text prompt in a certain way and in 4/4. Conversely, it might be easier to try to catch if an audio sample is the output of an AE. For instance, it is well-known that neural decoders tend to produce \textit{checkerboard artefacts} \cite{kumar2019melgan} characteristic of transposed convolution operations. We might be able to catch many more such artefacts. Thus, we propose the following research direction: \textbf{can we detect if a music sample is generated by an artificial decoder, this independently of its musical content?}

Another difficulty is of a causal nature. If we collected real and fake music samples and naively trained a model to classify them, we might end up detecting features unrelated to generation artefacts. For instance, a public real music dataset might be full of classical music, while a deepfake dataset primarily includes rap and pop music. This could result in the classifier learning to detect classical music instead of distinguishing real and forgeries. This problem is known as \textit{confounding} \cite{peters2017elements}.
The same discussion applies to the compression codec that might confound the detection of deepfakes (\eg all Riffusion songs are exported in mp3 192kB/s).

These two remarks have motivated our following framework. We leverage a dataset of real music samples, which we auto-encode using the trained AE part of the above models. These samples are stored at the same bitrate as the original audio. Controlling on the music semantic and file encoding, the model we then train can only detect generation artefact since it should learn to tell apart a real audio from its reconstructed counterpart. Therefore, we avoid any extraneous confounding influence.

\subsection{Considered dataset and deepfake generators}

We chose to use the FMA dataset \cite{fma_challenge}, an open dataset that allows reproducibility and comparison of future work. Due to size constraints, we only consider the medium split that includes 25.000 music tracks spread into 16 genres. All tracks are encoded in mp3 with a diversity of bitrates -- with a majority of 320kB/s, followed by 256 and 192kB/s. We had to resample a few tracks to 44.1kHz for consistency.

As for the autoencoders, we consider two popular neural codecs: Encodec (\eg used by \textit{Suno v2} and MusicGen) and DAC (\eg used by Vampnet\footnote{For interested readers, we actually use the LAC version of DAC that is better suited for music, similar to what is done in VampNet \cite{garcia2023vampnet}.}). We also studied the decoder part of Musika, which was trained end-to-end on polar spectrograms. Finally, we consider a combination of a mel-spectrogram converter-inverter and a Griffin-Lim phase reconstruction (\eg used by \textit{Riffusion v1}) -- we dub this pipeline \textit{GriffinMel}. Some audio reconstructions are available on our repository to gain intuition on each decoder.
The availability of trained models constrained our choice of decoder. For instance, Soundstream is used as a latent representation in many of Google's models, yet no public checkpoint is available. The same applies to MelGAN and HiFiGan, for which no checkpoint exists for music data\footnote{Although there exists some for voice synthesis, we found that this resulted in heavy audible artefact on music that we deemed unrealistic.}, as well as \textit{Riffusion v3} and \textit{Suno v3} that are now closed-source.
We consider several configurations for the above decoders: Encodec in 3, 6 and 24kB/s, DAC in 2, 7 and 14kB/S, and GriffinMel using 256 and 512 melbands.

We autoencode all considered real tracks and obtain 249820 tracks spread across a real class and nine synthetic reconstructions. We split the audios in a 70\%, 10\%, 20\% fashion between train, validation, and test.
Empirically, the GriffinMel reconstruction seems the easiest to catch due to audible phase errors.
Encodec and DAC sound most challenging, especially at their maximum bitrate. If it is often possible to distinguish between a real and a fake when placing one next to the other, it is way more tricky without a point of reference or being aware that the audio could be fake. This relates to the recent user study in \cite{cooke2024good}.

\subsection{First results}

We started experimenting with our dataset with straightforward convolutional models.
We use a standard optimiser setting (\ie Adam with progressive learning rate decrease). During training, real and fake music tracks are sampled with a $\frac 1 2$ probability, then, the fake is sampled uniformly among the nine decoders.
Finally, we extract random 0.8s from each track of the batch.
To our surprise, this basic setting led to test accuracies over 90\%, which we were initially sceptical about. After several experiments, it seems this detection task is easier than we thought. The choice of architecture does not seem to impact the performance much. However, the choice of input preprocessing seems to be much more influential. We report in Table \ref{table:results} the detection score for various choices of audio representations: the raw waveform, the complex STFT, its amplitude, its phase, or both stacked as polar coordinates. All our results are very satisfying overall.
Transforming music samples as \textit{amplitude} spectrograms leads to the best performance overall. It is also interesting to see that the purely \textit{phase}-based model yields high scores despite often being considered less efficient than the amplitude representation.

\begin{table*}[h]
    \centering
    \includegraphics[height=6cm, trim={0 0.25cm 0 0.5cm},clip]{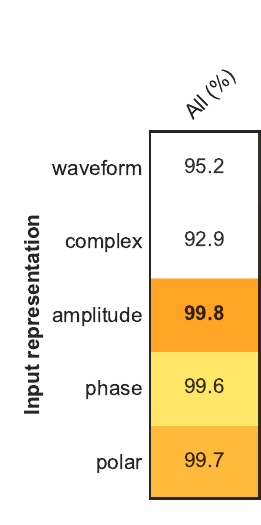}
    \hspace{0.25em}
    \includegraphics[height=6cm, trim={0 0.25cm 0 0.5cm},clip]{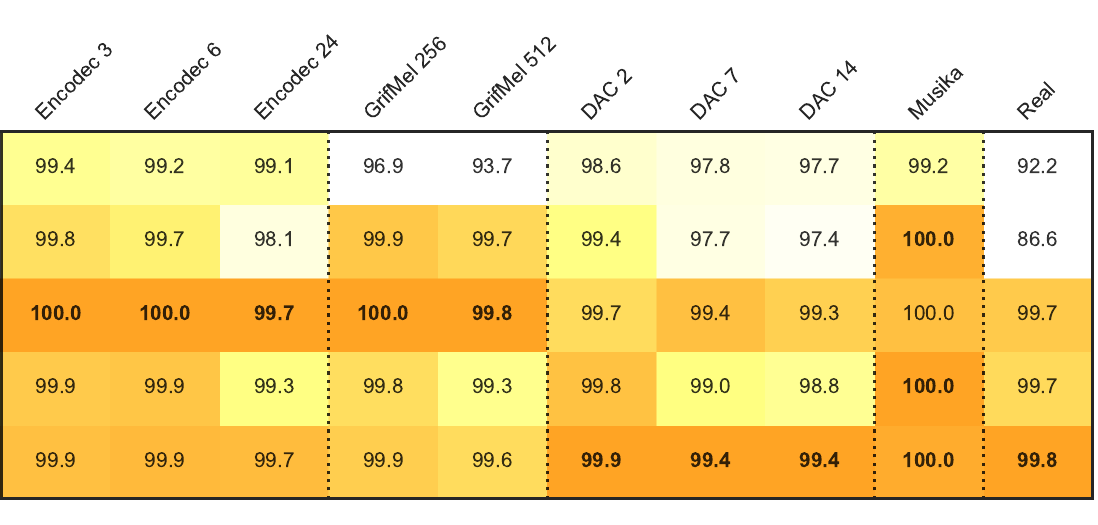}
    \caption{\textbf{Test accuracy detection scores (\%) for different audio representations}. We include a per-class breakdown.
    The label numbers indicate the varying parameter configuration specific to each method, if applicable.
    Background colours highlight the \textit{best scores per column}. In the rest of the paper, we chose the \textit{amplitude}-spectrogram-based model as the best model for our experiments.}
    \label{table:results}
\end{table*}

For conciseness, we do not report results on architectural changes.
It is possible that the above ranking of best audio representation might change with a different configuration than what we tested.
Briefly, our proposed model is composed of six convolutional layers with $[16, 32, 64, 128, 256, 512]$ filters. We use kernels of size 3 and a pooling of size 2. We finish with an average pooling and two linear layers. This amounts to 1.6M trainable parameters, which is small relative to the generative models we study. 
We use some light preprocessing of the audio (\eg normalisation, random mono mix, cutoff at 16kHz, and conversion to decibel scale when applicable).
All details and trained weights are available on our repository.

We underscore that given our resulting performances, we did not find it so crucial to explore the best possible architecture further and deemed it more important to discuss the aftermath of obtaining such convincing scores. Indeed, it could feel that we have "solved" the task. Nevertheless, should we be so confident?

\section{Caveats on deepfake detectors}

We find it crucial to take a step back from a purely academic ML perspective and ponder the larger consequence of deploying a deepfake detector.
Indeed, a new market of "AI content detectors" has emerged in recent years: \eg checking if a student essay employed ChatGPT\footnote{\eg \url{https://gptzero.me/}}. Such tools often claim to have high detection scores. However, they are often closed-source, making verification tricky.
This has notably led to strange situations where students have much trouble proving their good faith in false positive cases against the ethos of a so-called "trusted AI checker" \cite{url_ai_detection}.
This section hence discusses aspects to make deepfake detection more reliable and ethical.
We open doors to four facets that make deepfake detection more complex than may first appear: robustness to audio manipulation, generalisation to unknown encoders, calibration discussion, possibility for recourse and interpretability.

\subsection{Robustness to manipulations}

An angle often discussed in the literature on forgery detection is the robustness to data shifts.
There are countless scenarios where deepfake creators do not directly publish the immediate output of the generative model. For instance, they could genuinely reencode them in a different format while exporting the result or adding it to a video clip. They could also try to bypass a detector more strategically by applying time-stretching or pitch shift transforms, similar to what is frequently done on social networks to bypass fingerprint systems and evade copyright claims.
It would be unrealistic not to expect some users to try to evade detection.
However, this is typically a cat-and-mouse game, where it's illusory to anticipate all attacks in advance. Attackers will always find new ways to circumvent the countermeasures.
We believe this is a much more continual process of making it gradually difficult to evade a detector by patching it regularly (\ie similar to an antivirus software). Therefore, evaluating detection models in a scenario of partial knowledge is essential.

As a first study, we consider some common audio transformations that lay users could employ: random pitch shift ($\pm2$ semitones), time stretch ($[80,120]\%$), EQ, reverb, addition of white noise, reencoding in \textit{mp3}, \textit{aac}, and \textit{opus} in 64kB/s.
Implementation details are available in our repository.
We leave attacks from more advanced users for future work (\eg adversarial attacks \cite{szegedy2013intriguing}). We evaluate the best model from the previous section on such unseen transformation and report the results in Table \ref{table:robustness}.
The performances drop drastically under pitch shifts, the addition of white noise, and codec reencoding. This is consistent with previous literature on forgery detection that ML models are generally not robust to out-of-distribution shifts -- if not explicitly designed for them \cite{wang2020cnn, li2020celeb}.
Conversely, it is unclear why the model remains robust to some manipulations (\ie time stretch, EQ and reverberation).

We highlight that several scores drop to almost zero, which means that the model has predicted the real class for most samples (instead of more unconfident, aleatoric guesses). Meanwhile, the manipulations did not impact the \textit{real} class scores. 
This suggests that the model works by detecting artefacts specific to each AE and otherwise defaulting to the \textit{real} class if none is found (or that the manipulations make them unrecognisable for the model). This would not be surprising since the \textit{real} class can be expected to be more diverse and complex than autoencoded generations \cite{tishby2000information}. Since ML models are biased toward simple solutions \cite{scimeca2021shortcut}, it is expected that it is easier to detect a real audio by \textit{not detecting the learned characteristics of fakes} instead of learning a manifold of real music.
This is a critical remark because we can already anticipate that the model may not generalise to unseen deepfake generators.

Note that we did not train the models on said manipulations, \ie data augmentation. We merely checked their natural robustness.
In some subsequent experiments, we saw that fine-tuning on these manipulations could reliably restore high accuracy scores. However, in this paper, we prefer to hammer home the following: \textit{there will always be an unseen manipulation}. It would not be realistic to only evaluate our model on manipulations we optimise for. Overall, our results suggest that deepfake detectors are not naturally robust to unanticipated audio transformations.

\begin{table*}[h]
    \centering
    \includegraphics[height=6.75cm, trim={0 0.25cm 0 0.25cm},clip]{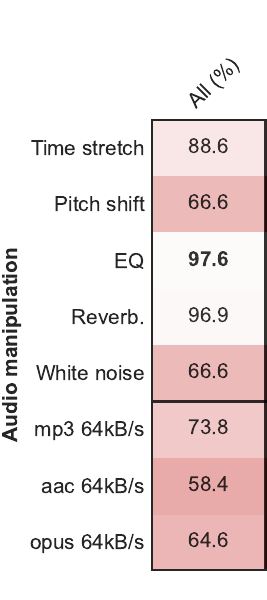}
    \hspace{0.25em}
    \includegraphics[height=6.75cm, trim={0 0.25cm 0 0.25cm},clip]{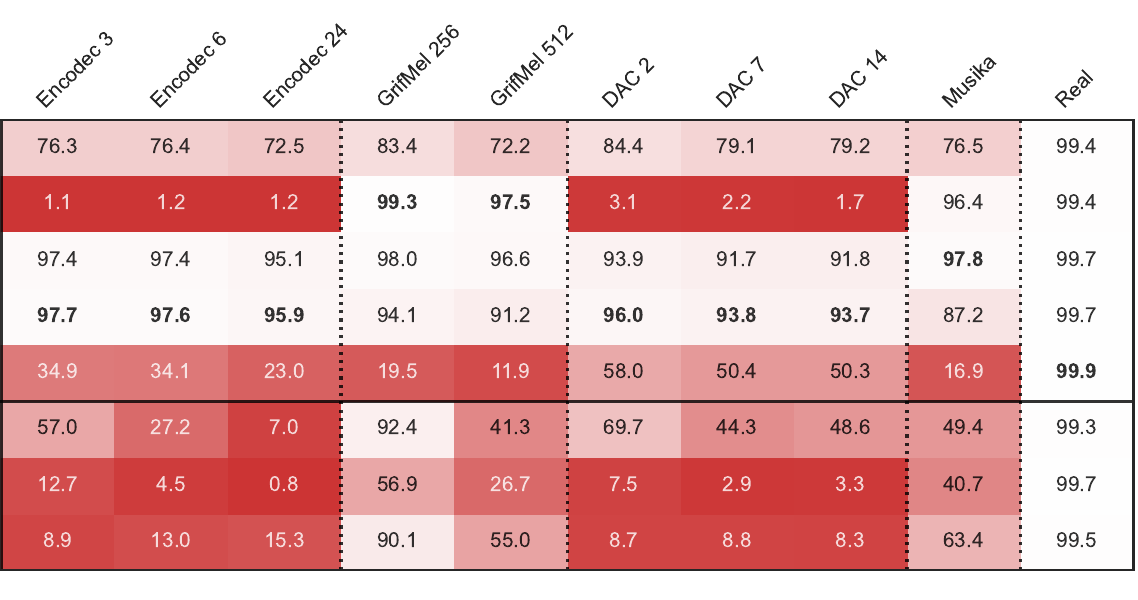}
    \caption{\textbf{Robustness accuracy test scores (\%) of the amplitude-spectrogram-based model to audio transforms}. We include a breakdown per class as in Table \ref{table:results}.
    Background colours highlight \textit{strong score degradation}.
    }
    \label{table:robustness}
\end{table*}

\subsection{Encoder generalisation}

Another important question is whether our detector generalises to AE models that were not considered during training. Instead of finding additional AE to test, we conduct a new experiment in which we retrain our best model from scratch on each of the nine considered decoders (versus real audios) and check how the detection performance naturally transfers to the others. The results are displayed in Table \ref{table:generalisation}.
Interestingly, we first find that the models are pretty robust \textit{intra}-family: \eg learning on Encodec 24kB/s reconstruction transfers well to 6kB/s and 3kB/s. It is reassuring that we may not need to include all possible parametrisation of an AE to learn to detect it.
Learning from a higher bitrate seems to transfer better to low bitrate, which could stem from the RVQ formulation of the considered models, but this is not so straightforward to assert.

Then, we note that the model falters on \textit{inter}-family generalisation: said performances are almost always zero (\eg GriffinMel $\rightarrow$ DAC).
This aligns with the previous section that the models are not robust to unseen manipulations.
Note that the performances drop again to 0\%, which implies that the \textit{real} class may be acting as a default.

\begin{table}[h]
    \centering
    \includegraphics[width=\linewidth]{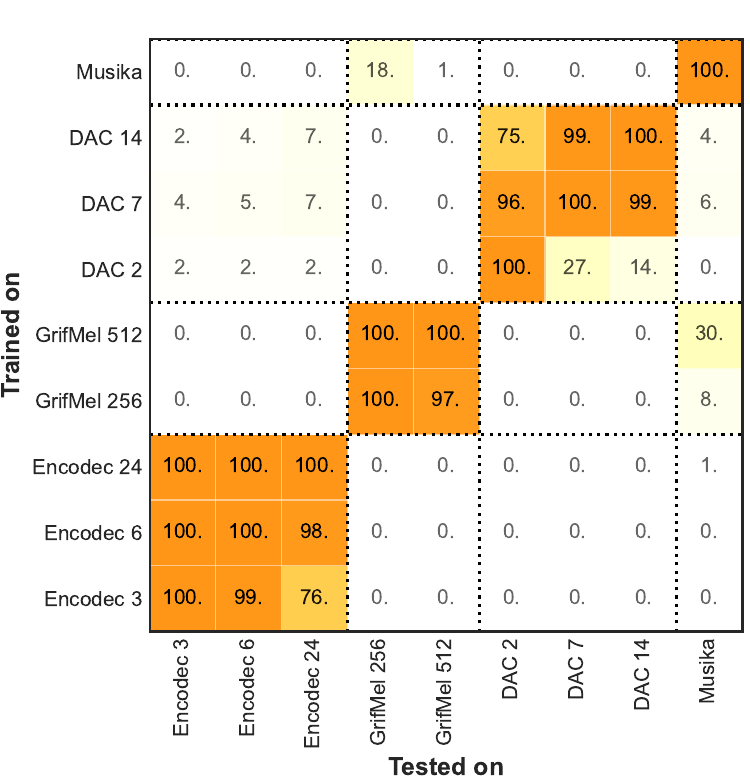}
    \caption{\textbf{Generalisation to unseen deepfake generators}. We train on each single decoder indicated on the left axis and evaluate on each test accuracy (\%) of the bottom axis.
    }
    \label{table:generalisation}
\end{table}

\subsection{Calibration and specifications}

Our previous remark on the real class acting as default is a first step toward thinking about the \textit{technical specifications} of a detector to be deployed to the general public.
As mentioned before, if it is legitimate to combat AI by AI, we also need to underscore that many of the risks posed by generative AI also apply to AI detectors.
Returning to the parallel case of AI-text checkers, one big impediment to trusting these models is that they are typically miscalibrated. In fact, most ML models are miscalibrated, and overconfident \cite{nguyen2015deep, guo2017calibration}.
In our music case, this would make it hard for artists to argue against a claim that their music is estimated to be "99\% fake" (\ie false positive), if the involved stakeholders are unaware of this miscalibration aspect, this even more so with closed-source technology.
We must thus ensure that deepfake detectors output probabilities that make sense.

\begin{figure}[h]
    \centering
    \includegraphics[width=0.875\linewidth]{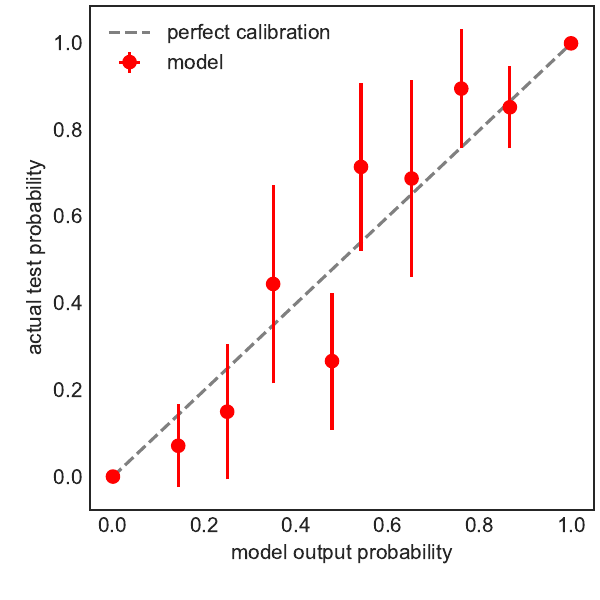}
    \caption{\textbf{Model calibration curve}. Note that predictions in $[0.1, 0.9]$ are in minority, which results in higher confidence intervals at 95\% in that range.}
    \label{fig:calibration_1}
\end{figure}

This is yet again a vast topic, and we merely wanted to open some doors to it. As a first step, we can compute the calibration curve of our best model, namely, compare the output probability to effective performance. If the model is well calibrated, predicting a 70\%-80\% probability should accordingly lead to a 70\%-80\% accuracy conditioned on all predictions of said range. We display the calibration curve in Figure \ref{fig:calibration_1}. Contrary to our expectations, the model actually looks rather calibrated although we did not take any measure for it. However, a large majority of predictions are very close to 0 or 1 with the dataset we use and lead to large confidence intervals for intermediate values.

\vspace{0.8em}

We wanted to push this discussion further and ran a second experiment to study said "intermediate" values in more detail with more diverse music samples that could be encountered in the wild.
Indeed, beyond the vanilla calibration issues we have just discussed, another practical type of music deepfake may be ones where only some stems of the music are fake: \eg a fake singing voice over real instrumentals, or the other way round with a rap over a generated backing track. In that case, what score should a detector model display? 100\% fakeness due to the presence of any forgery in the track, or, some \textit{fakeness ratio}? What solution does the model naturally fall into (if any)?

\begin{figure}[h]
    \centering
    \includegraphics[width=0.875\linewidth]{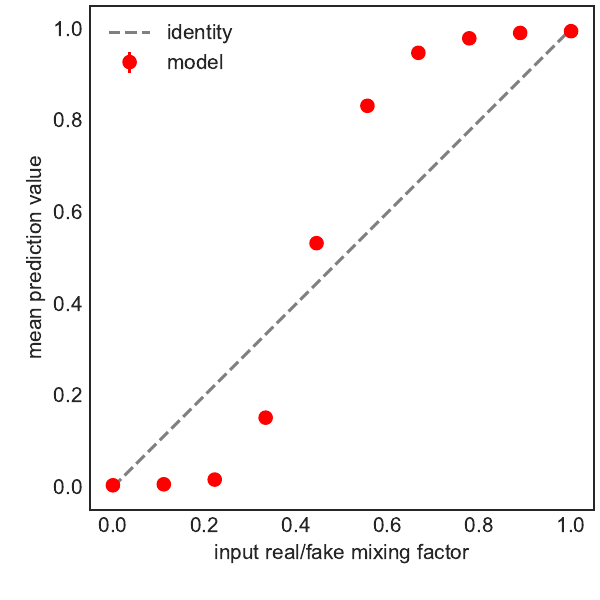}
    \caption{\textbf{Mixing calibration curve}. We generate an equal number of real/fake mixes (4955) per mixing factor. The confidence intervals are too small to be visible here.}
    \label{fig:calibration_2}
\end{figure}

Instead of mixing real and fake stems, as an opening experiment, we compute prediction on music audios that mix real tracks and fake counterparts with a gradual fading coefficient, which is straightforward to generate with our dataset (and the audios are aligned). The calibration curve is given in Figure \ref{fig:calibration_2}. Arguably, the model follows the first solution with a switch at 50\%, instead of a solution involving the true mixing ratio. However, it was not particularly designed for either. This is thus not a "failure" but an insight into the model's functioning.
Of course, this experiment is not realistic of a true stem mixing and audio engineering. We rather envision it as a discussion starter on the topic.
Our take is that there is not necessarily a "best" expected behaviour that the detector should adhere to for this type of content. Researchers, engineers, product owners, and regulators should all be involved in the model specification. Furthermore, this sort of curve could be made accessible to the general public to help interpret a detector's output.

\subsection{Recourse and interpretation}

Ensuring model calibration is essential, but how do we manage actual false positive cases? Arguably, there should be a way to check why a prediction was made and address errors manually. This angle relates to interpretability and recourse.

As a short proof of concept, we show that we can compute some \textit{feature attribution} maps of what part of the spectrogram is considered fake. This is useful to address the earlier case of a fake voice sung over real music.
Feature attribution is a technique for explainability aiming to relate the influence of an input on an output. However, the exact specification of the posed question the attribution explanation answers to is not always so clear to spell out and can lead to misunderstanding from the recipient of the explanations \cite{lombrozo2006structure, adebayo2022post, miller2023xaiisdead}.
Here, we assume that the input could contain some mixed fake and real stem and that an amplitude spectrogram is a good explanation space to highlight such an effect. By comparison, this would not be the case if the fakeness of a spectrogram was related to its overall bluriness: a feature attribution map would not help adduce a reason for the prediction since the whole (blurry) spectrogram would be highlighted.
The latter case would rather call for a change of the explanation space (\eg concept learning \cite{kim2018interpretability, afchar2022learning, foscarin2022concept}).
Therefore, our following experiment only applies to explaining mixed stems music but does not holistically address interpretability.

\begin{figure}[h]
    \centering
    \includegraphics[width=\linewidth]{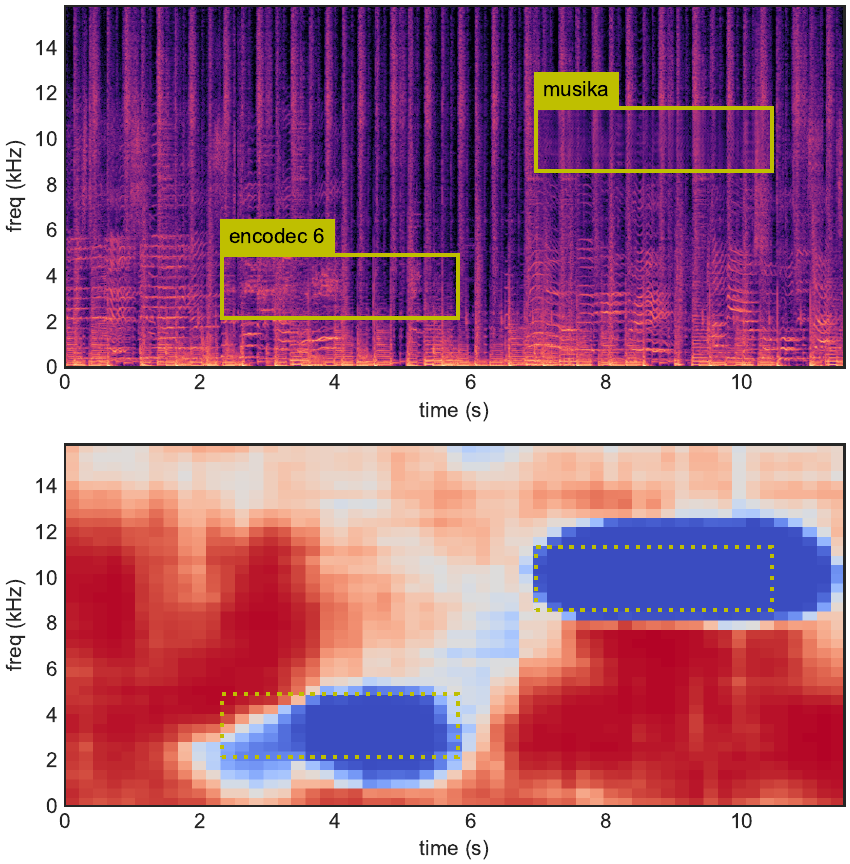}
    \caption{\textbf{Feature attribution example}. Mixed real/fake spectrogram on the top, corresponding per-patch predictions on the bottom (red for \textit{real}, blue for \textit{fake}).  }
    \label{fig:interpretability}
\end{figure}

Since we leverage an average pooling in our model, we can change its input size and
get more localised predictions.
We fine-tune our model on the smallest possible size (\ie the receptive field), on randomly cropped spectrogram patches.
Then, iterating predictions over a whole spectrogram leads to the feature attribution we were looking for.
Figure \ref{fig:interpretability} provides an application example: a real music spectrogram has two patches replaced with some auto-encoded counterparts, hence simulating a partially fake signal.
Our proposed computation of feature attribution maps can help localise this type of manipulation.

\vspace{0.25em}
Nevertheless, let us acknowledge that the scope of the previous discussion should not be restricted to purely technical insights.
Fighting deepfakes with an AI deepfake detector is a form of techno-solutionism.
It can lead to a myopic view of the topic, potentially overlooking other parts of the deepfake supply chain beyond ML's rationale \cite{parliament2021, miotti2024combatting}. For instance, it might be more efficient to regulate big tech actors from training music generators on millions of tracks\footnote{ofttimes copyrighted, which is another issue} and making them available to the world, than putting off fires of detecting these generations afterward. An alternative lead could be to watermark generations \cite{chen2023wavmark}, preventing the bulk of lay users from spreading unlicensed generations. Alternatively, to enforce strict policies on music generation, or leave the job to musicians, as many artists call (\eg \cite{gebru2023artists}, ARA\footnote{\url{https://artistrightsalliance.org}}). However, the latter does not seem likely to happen.
Interpretability can similarly turn bad if only addressed mathematically and with the illusion of objectivity \cite{crawford2021atlas}. There is a common \textit{dilemma of control} for tech companies that fear giving more agency to users, while interpretability also starts when the power balance shifts towards more stakeholders \cite{hardt2022performative, afchar2023interpretable}. We need a more encompassing political vision about the models we deploy \cite{green2021data, parliament2021}.
In a nutshell, while interpretability is crucial, it is important to acknowledge that (in this case) its use is a bandage on an already bad situation when clear regulations on generative technology could address many aspects of the topic more efficiently.

\section{Conclusion}

In this paper, we proposed the first study on music deepfake detection. We show that such forged content is surprisingly easy to detect, yet stress that a good accuracy score is not at all the end of the story. Therefore, we recommend considering (at least) four additional aspects: robustness to manipulation, generalisation to unseen generators, calibration and interpretability.

An expected counter-argument to our work is whether employing a bigger model (\eg ResNet as in \cite{zang2024singfake}) would solve everything. We are very sceptical of this stance and instead urge to integrate explicitly designed objectives instead of hoping these crucial problems sort themselves out. Our exposed caveats should not be seen as desirable by-products of training but rather as must-have features and discussed specification before any deployment of a reliable and ethical detector.

Our future work includes studying whether these models can be easily fine-tuned or updated for new generators, their generalisation capabilities with further data augmentation during training (\eg audio manipulations), defense against adversarial attacks, and the impact of more realistic stem mixing and audio engineering.

\section{Ethics statement}

We have already evoked a number of ethical risks posed by the elaboration of deepfake detectors. In the light of existing AI-content checkers, we know that false positive cases have led to many hairy situations. Of course, this can be amplified by marketing hype around a given detector that makes it look more reliable than it really is, but also by a lack of interpretability or the closed source of said tool.
This also relates to the vicious circle of techno-solutionism, when the only options considered to regulate deepfakes are to create more AI products. As we mentioned, our technical results are only one piece of a larger puzzle involving many stakeholders, and whose completion should not be left entirely to the decision of a few private companies.

In this paper, we have also chosen the term "fake" to refer to artificial generations. This may feel like it evacuates the possibility of a fair use of generative AI, \ie a new musical means of expression and somewhat new synthetic instruments.
In this context, detectors might only be necessary for actors not complying with established rules (\eg watermarking their generations).
While we agree that an ethical use of this technology might be possible, it would be naive to overlook that most current generative technologies are not going in this direction.
Regrettably, the current norm instead seems to be predatory uses of AI technology without any consent, fair compensation or credit for the pool of artists that enabled this technology to exist in the first place.
Or rather, this approach is currently the most financially rewarding.
This is why, in this paper and at this moment in time, we deemed it important to frame the topic as a risk instead of a more neutral, decontextualised, technical advance.

\bibliography{bib}

\end{document}